\input{aipcheck}

\documentclass[
    ,final            
  ]
  {aipproc}

\layoutstyle{8x11double}

\begin{document}

\title{Stochastic Electron Acceleration in Shell-Type Supernova Remnants II}


\author{Siming Liu}{address={Los Alamos National Laboratory, Los Alamos, NM 87545} }
\author{Zhong-Hui Fan}{address={Department of Physics, Yunnan University, Kunming 650091, Yunnan, China; zhfan@ihep.ac.cn} }
\author{Christopher L. Fryer}{address={Los Alamos National Laboratory, Los Alamos, NM 87545} }

\begin{abstract}
We discuss the generic characteristics of stochastic particle
acceleration by a fully developed turbulence spectrum and show that
resonant interactions of particles with high speed waves dominate
the acceleration process. To produce the relativistic electrons
inferred from the broadband spectrum of a few well-observed
shell-type supernova remnants in the leptonic scenario for the TeV
emission, fast mode waves must be excited effectively in the
downstream and dominate the turbulence in the subsonic phase. Strong
collisionless non-relativistic astrophysical shocks are studied with
the assumption of a constant Aflv\'{e}n speed. The energy density of
non-thermal electrons is found to be comparable to that of the
magnetic field. With reasonable parameters, the model explains
observations of shell-type supernova remnants. More detailed studies
are warranted to better understand the nature of supernova shocks.
\end{abstract}

\maketitle


\section{Turbulence Cascade and Stochastic Particle Acceleration}
\label{TS}

In the Kolmogorov phenomenology, the free energy dissipation rate is
given by $Q \equiv C_1\rho u^3/L\,,$ where $C_1$ is a dimensionless
constant, and $u$ and $L$ are the eddy speed and the turbulence
generation scale, respectively.  The eddy turnover speed and time at
smaller scales are given respectively by $v^2_{edd}(k)\equiv4\pi
W(k)k^3\propto k^{-2/3}$ and $\tau_{edd}(k) \equiv2\pi /kv_{edd}=
\pi^{1/2}(W k^5)^{-1/2}\propto k^{-2/3}\,,$ where $W(k) =
(u^2/4\pi)(2\pi/L)^{2/3} k^{-11/3}= (4\pi)^{-1} (2\pi
Q/C_1\rho)^{2/3} k^{-11/3}\propto k^{-11/3}\,$ is the isotropic
turbulence power spectrum, $k=2\pi/l$ is the wave number and $l$ is
the eddy size. From the three-dimensional Kolmogorov constant $C
\simeq 1.62$ \citep{yz97}, we obtain $C_1 =  2\pi/C^{3/2} = 3.05\,.$
At the turbulence generation scale $k_m = 2\pi/L$, $v_{edd}=u$, $Q=2
C_1\rho [4\pi W^3k^{11}]^{1/2}= C_1\rho v^2_{edd}(k)/
\tau_{edd}(k)\,,$ and the total turbulence energy is given by $\int
W(k) 4\pi k^2 {\rm d} k = (3/2) u^2\,.$ The turbulence decay time is
therefore given by $\tau_d = 3\tau_{edd}(k_m)/C_1\,,$ i.e., eddies
decay after making $3C_1^{-1}\sim 1$ turns.

We are interested in the acceleration of particles through
scattering randomly with heavy scattering centers with the
corresponding acceleration time $\tau_{ac} = \tau_{sc} [3
v^2/v^2_{edd}(k)]\,,$ where $\tau_{sc}= 2\pi/kv=l/v$ is the
scattering time, $v$ is the particle speed, and we have assumed that
the scattering mean free-path is equal to $l$. For the above
isotropic Kolmogorov turbulence spectrum, $\tau_{ac}(k) = 3
v/2Wk^4\propto k^{-1/3}\,.$ To have significant stochastic particle
acceleration (SA), the acceleration time $\tau_{ac}(k)$ should be
shorter than the turbulence decay time $\tau_d$, which implies
$u^2>C_1 v v_{edd}(k)\,.$ So, in general, the SA is more efficient
at smaller scales. The onset scale of the SA is given by $k_c =
(C_1v/u)^3 k_m\,.$ Therefore, to produce energetic particles with a
speed of $v$ by a Kolmogorov spectrum of scattering centers, the
turbulence must have a dynamical range greater than
$D_k=(C_1v/u)^3\,.$

In the Kraichnan phenomenology, the turbulence decay is suppressed
by the wave propagating effect with
$Q=C_1\rho u^4/Lv_F,$
$W(k) = (u^2/4\pi)k_m^{1/2}k^{-7/2},$
$\tau_{edd}\propto k^{-3/4}\,, v_{edd}\propto k^{-1/4}\,,
\tau_{ac}\propto k^{-1/2}\,,$ and the turbulence decay time
$\tau_d=3\tau_{edd}(k_m)v_F/C_1u\,,$ where the wave speed $v_F\gg
u$. To have significant acceleration through scattering
with the eddies, the dynamical range of the turbulence must be
greater than $D_w=(C_1v/v_F)^2\,,$ which is much less than
$D_k=(C_1v/u)^3$.
The resonant interactions of particles with waves can be much more
efficient in accelerating particles in this case. For a wave speed
$v_F$ independent of $k$, the acceleration time is given by
$\tau_{ac}=3 \tau_{sc} v^2/v_F^2\propto k^{-1}\,.$ To have
significant acceleration, the scattering mean free path of the
particles must be shorter than $(v_F^3/C_1vu^2)L= D_k^{2/3}
D_w^{-3/2}L\,.$

Several Shell-Type Supernova Remnants (STSNRs) have been observed
extensively in the radio, X-ray, and TeV bands. X-ray observations
with {\it Chandra}, {\it XMM-Newton}, and {\it Suzaku} and TeV
observations with HESS have made several surprising discoveries that
challenge the classical diffusive shock particle acceleration model
\citep{l08,t08}. The SNR RX J1713.7-3946 is about $t =1600$ years
old \citep{w97} with a radius of $R\simeq 10$ pc and a distance of
$D \simeq 1$ kpc.  By fitting its broadband spectrum with an
electron distribution of $f\propto \gamma^{-p} \exp
-(\gamma/\gamma_c)^{1/2}$, we find that $p = 1.85$, $B = 12.0\;
\mu$G, $ \gamma_c m_e c^2 = 3.68$ TeV, and the total energy of
relativistic electrons with the Lorentz factor $\gamma>1800$
$E_e=3.92 \times 10^{47}$ erg (Fig. \ref{spec}).

The X-ray emitting electrons have a gyro-radius of $r_g\simeq
10^{15}$ cm, which shouldn't be shorter than the scattering mean
free path. To produce these electrons through the SA, the turbulence
must be generated on scales greater than $D_kr_g$, $D_wr_g$, and
$D_w^{3/2}D_k^{-2/3}r_g$ for the non-resonance Kolmogorov, Kraichnan
phenomenology, and the resonant interactions, respectively. For
STSNRs, $u\sim v_F\sim 0.01c$, $D_kr_g\sim 10$ kpc, which is much
larger than the radii of the remnants. The SA by eddies with a
Kolmogorov spectrum is therefore insignificant. $D_wr_g\sim 30$ pc,
which is also too thick. $D_w^{3/2}D_k^{-2/3}r_g\sim 0.1$ pc, which
is much greater than the particle inertial length and may be
generated through the Kelvin-Helmholtz instabilities or cosmic ray
drifting upstream \citep{m99,n08}. Therefore if relativistic
electrons from the STSNRs are accelerated through the SA, they must
be energized through resonant interactions with high speed plasma
waves. Low speed waves also require a large turbulence dynamical
range to accelerate particles.

\section{Shock Structure, Wave Damping, and Stochastic Electron Acceleration by Fast Mode Waves in the Downstream}
\label{shock}

We next study the SA in the shock downstream by weakly magnetized
turbulence with the Alfv\'{e}n speed $v_A=(B^2/4\pi\rho)^{1/2}\ll
u$, where $B$, and $\rho$ are the magnetic field, and mass density,
respectively. For strong non-relativistic shocks with the shock
frame upstream speed $U$ much higher than the speed of the parallel
propagating fast mode waves in the upstream $v_F =
(v_A^2+5v_S^2/3)^{1/2}$, where $v_S^2=P/\rho$ is the isothermal
sound speed and $P$ is the gas pressure, mass, momentum, and energy
conservation across the shock front require
\begin{equation}
U^2= 5 v_S^2 + 5 u^2+ 2v_A^2+U^2/16\,, \label{energy}
\end{equation}
where we have assumed that the turbulence behaves as an ideal gas
and ignored the wave propagation effects. The shock structure can be
complicated due to the present of turbulence. We assume that the
turbulence is isotropic and has a generation scale of $L$, which
does not change in the downstream. The speeds $v_S$, $v_A$, and $u$
therefore should be considered as averaged quantities on the scale
$L$. $v_A$ depends on the upstream conditions and/or the dynamo
process of magnetic field amplification\citep{c00,n08}. Here we
assume it a constant in the downstream. One can then quantify the
evolution of other speeds in the downstream.

For the Kolmogorov phenomenology,
\begin{equation}
{3{\rm d} \rho u^2 \over 2{\rm d} t}= -Q\ \ \ {\rm i.e.,}\ \ \
{3U{\rm d} u(x)^2 \over 8{\rm d} x}  = -{C_1 u(x)^3\over L}\,.
\label{ux}
\end{equation}
Near the shock front, we denote the isothermal sound speed and
Aflv\'{e}n speed by $v_{S0}$ and $v_{A0}$, respectively. Then the
eddy speed at the shock front is given by $a^{1/2}U/4$ with $a = 3 - 16
v_{S0}^2/U^2 - 32 v_{A0}^2/5U^2$. Integrate equation (\ref{ux}) from
the shock front $(x=0)$ to downstream ($x>0$), we then have
\begin{eqnarray}
{u(x)\over U} &=& {1\over 4C_1x/3L+ 4/a^{1/2}}\,, \\
{v_S(x)\over U} &=& \left[{3\over 16}-{1\over
16\left(C_1x/3L+a^{-1/2}\right)^2}- {2v_A^2\over 5U^2}
\right]^{1/2}\,, \\
{v_F(x)\over U}&=& \left[{5\over 16}-{5\over
48\left(C_1x/3L+a^{-1/2}\right)^2}+ {v_A^2\over 3U^2} \right]^{1/2}\,.
\end{eqnarray}
\begin{figure}[t]
\includegraphics[height=.27\textheight,width=0.48\textwidth]{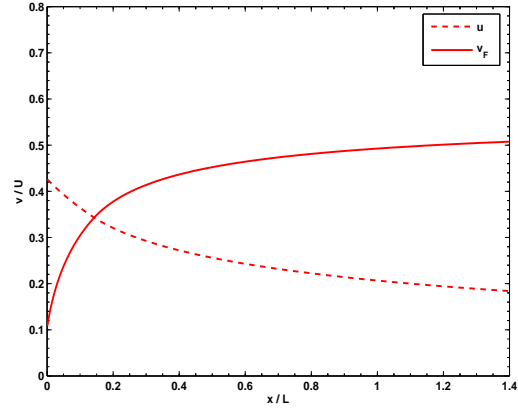}
\caption{Evolution of the eddy speed $u$ and speed of parallel
propagating fast mode waves $v_F$ in the downstream for $v_{A}=0.0633U$. \label{vs}}
\end{figure}
As mentioned in the previous section, to produce the observed X-ray
emitting electrons in the STSNRs through the SA processes, fast mode
waves must be excited efficiently. The MHD wave period is given by
$\tau_F(k)=2\pi /v_Fk$. Then the transition scale from the
Kolmogorov to Kraichnan phenomenology occurs at $\tau_F(k_t) =
\tau_{edd}(k_t)$ or $v_F = v_{edd}(k_t)$\citep{j08}. We then have
\begin{equation}
k_t = (u/v_F)^3k_m\,.
\end{equation}
For $k>k_t>k_m$, the turbulence spectrum in the inertial range is
given by $
W(k) 
= u^2(4\pi)^{-1}k_m^{2/3} k_t^{-1/6}k^{-7/2}=(4\pi)^{-1}
v_F^{1/2}u^{3/2}k_m^{1/2}k^{-7/2}\,.
$
Although the turbulence energy exceeds $(3/2)u^2$ when the wave
propagation effect is considered, we still assume that the
enthalpy of the turbulence is given by $(5/2) u^2$ for $v_F<u$ so
that equation (\ref{energy}) and the above solutions for the speed
profiles remain valid.

In the subsonic phase with $v_F>u$,  we assume that fast mode waves
can still be excited efficiently to maximize the SA efficiency. Then
the Kraichnan phenomenology prevails and
\begin{eqnarray}
W(k) &=& u^2(4\pi)^{-1}k_m^{1/2}k^{-7/2}\,,\\
{3U{\rm d}u(x)^2\over 8{\rm d} x}&=& -{C_1 u(x)^4\over Lv_F}
\end{eqnarray}
where from equation (\ref{energy}) one has $v_F = \left[{5U^2/16}+
{v_A^2/3-{5u^2(x)/3}} \right]^{1/2}.$ These equations can be solved
numerically to get the speed profiles in the subsonic phase. Figure
\ref{vs} shows the $v_F$ and $u$ profiles with $v_A=v_{A0}=0.0633U$
in the downstream and $v_{S0}=v_{A0}\ll U$.

\begin{figure}[t]
\includegraphics[height=.27\textheight,width=0.48\textwidth]{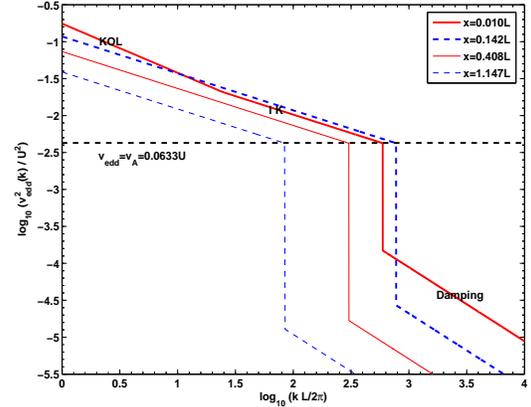}
\caption{The turbulence spectra $v_{edd}^2(k)$ at several locations
in the downstream as indicated. The Kolmogorov, Kraichnan, and
damping ranges are indicated for the supersonic phase spectrum with
$x=0.010L$. At the other locations, the turbulence is subsonic and
there are only Kraichnan and damping ranges. The sharp drops of the
turbulence spectra in the damping range are due to the onset of
thermal damping at the coherent length of the magnetic field
$2\pi/k_d$. \label{vedd}}
\end{figure}

In summary,
\begin{equation}
W(k)=u^{3/2}(4\pi)^{-1}\min({v_F^{1/2}, u^{1/2}})k_m^{1/2} k^{-7/2}\,
\label{IK}
\end{equation}
in the Kraichnan regime. The collisionless damping starts at the
coherent length of the magnetic field $ l_d = 2\pi/k_d\,, $ where
the period of Alfv\'{e}n waves $2\pi/kv_A$ is comparable to the eddy
turnover time, i.e., $ v_A^2 =  4\pi W(k_d) k_d^3 = \min({v_F^{1/2},
u^{1/2}}) u^{3/2}k_m^{1/2} k_d^{-1/2}\,. $ Then we have
\begin{equation}
k_d = [u^{3}\min(v_F, u)/v_A^4] k_m\, .
\end{equation}
For a fully ionized hydrogen plasma with isotropic particle
distributions, which is reasonable in the absence of strong large
scale magnetic fields, the transit-time damping (TTD) rate is given
by \citep{s62, pyl06}
\begin{eqnarray}
&&\Lambda_T(\theta, k) = {(2\pi k_{\rm B})^{1/2}k\sin^2\theta\over
2(m_e+m_p)\cos\theta} \times \nonumber \\
&&\left[\left(T_em_e\right)^{1/2} e^{-{m_e\omega^2\over 2 k_{\rm B}
T_ek_{||}^2}} + (T_pm_p)^{1/2} e^{-{m_p\omega^2\over 2k_{\rm B}
T_pk_{||}^2}}\right] \label{d1}
\end{eqnarray}
where $k_{\rm B}$, $T_e$, $T_p$, $m_e$, $m_p$, $\theta$, $\omega$,
and $k_{||}=k\cos\theta$ are the Boltzmann constant, electron and
proton temperatures, masses,  angle between the wave propagation
direction and mean magnetic field, wave frequency, and parallel
component of the wave vector, respectively. The first and second
terms in the brackets on the right hand side correspond to damping
by electrons and protons,
respectively. 
For weakly magnetized plasma with $v_A<v_S$, proton heating always
dominates the TTD for $\omega^2/k_{||}^2\sim v_S^2\sim k_{\rm B}
T_p/m_p$. If $v_A$ does not change dramatically in the downstream,
the continuous heating of background particles through the TTD
processes makes $T_p\rightarrow (m_p/m_e) T_e$ since the heating
rates are proportional to $(mT)^{1/2}$, where $m$ and $T$ represent
the mass and temperature of the particles, respectively. We see that
parallel propagating waves (with $\sin \theta=0$) are not subject to
the TTD processes and can accelerate some particles to very high
energy through cyclotron resonances. Obliquely propagating waves are
damped efficiently by the background particles. Although the damping
rates for waves propagating nearly perpendicular to the magnetic
field ($\cos\theta \simeq 0$) are also low, these waves are subject
to damping by magnetic field wandering \citep{pyl06}. The turbulence
power spectrum cuts off sharply when the damping rate becomes
comparable to the turbulence cascade rate $\Gamma =
\tau_{edd}^{-2}/(\tau^{-1}_F +\tau^{-1}_{edd})\simeq
\tau_{edd}^{-2}\tau_F$ \citep{j08}.  One can define a critical
propagation angle $\theta_c(k)$, where $\Lambda_T(\theta_c, k) =
\Gamma(k)$. Equations (\ref{IK}) and (\ref{d1}) give
\begin{equation}
{v_A^2 k_d^{1/2} \over 2^{1/2}\pi^{3/2}  v_S v_F k^{1/2}} 
\simeq  {\sin^2\theta_c\over \cos\theta_c} \exp\left(-{v_F^2 \over
2v_S^2\cos^2\theta_c}\right) \,,
\end{equation}
where the electron damping term has been ignored. The turbulence
spectra at several locations in the downstream are shown in Figure
\ref{vedd}.

\begin{figure}[t]
\includegraphics[height=.27\textheight,width=0.48\textwidth]{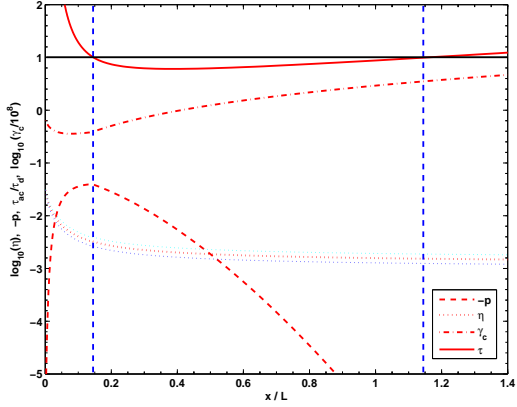}
\caption{Evolution of the acceleration efficiency $\eta$ (dotted),
cutoff Lorentz factor $\gamma_c$ (dotted-dashed), spectral index $p$
(dashed), and $\tau = \tau_{\rm ac}/\tau_{\rm d}$ (solid) in the
downstream for $v_{A}=0.0633U$ and $U=0.01c$. The particle
acceleration is significant for $\tau< 1$, i.e., between the two
vertical dashed lines indicating $x_1$ and $x_2$. For $\gamma_c$, we
have assumed that $B=12\mu$G and $L = 0.15$pc. For $\eta$, the two
thin lines are for $U=0.015 c$ (higher) and $0.0067c$ (lower). See
the following section for details \label{paras}}
\end{figure}

The escape time of relativistic particles with $v\simeq c$, where
$c$ is the speed of light, from the particle acceleration region is
given by $\tau_{esc} = (L^2/4c^2)\tau_{sc}$ and the spectral index
of the accelerated particles in the steady state is given by
\begin{eqnarray}
p & = & \left({9\over4} + {\tau_{ac}\over
\tau_{esc}}\right)^{1/2}-{1\over 2} = \left({9\over4} + {12c^2
k_m^2\over v_F^2 k_d^2}\right)^{1/2}-{1\over 2} \nonumber \\
& = & \left[{9\over 4} + {12c^2v_A^8 \over u^6 v_F^2 \min(v_F^2,
u^{2}) }\right]^{1/2}-{1\over 2}\,. \label{index}
\end{eqnarray}
We note that for $v_A$ independent of $x$ in the downstream, $p$
reaches its minimum at the transonic point $x_0$, where $v_F=u$. The
maximum energy that particles can reach though resonant interactions
with these parallel propagating waves is given by
\begin{equation}
\gamma_c={2\pi qB\over m_ec^2k_d}= {qBL v_A^4\over
m_ec^2u^3\min(v_F, u)}
\,, \label{cutoff}
\end{equation}
where $q$ is the elementary charge units. The ratio of the
dissipated energy carried by non-thermal particles to that of the
thermal particles should be greater than
\begin{equation}
\eta ={\theta_c^2(k_d)\over 4} ={e^{5/6}v_A^2\over 2(2\pi)^{3/2} v_S
v_F}= {e^{5/6}v_A^2\over 2(2\pi)^{3/2} v_S v_F}\,, \label{efficiency}
\end{equation}
where $e=2.72$, since the isotropic turbulence with $k<k_d$ can also
accelerate particles with the Lorentz factor $\gamma\ge \gamma_c$.
To have efficient acceleration of relativistic particles, the
turbulence decay time $\tau_d =3\max(u, v_F)L/C_1u^2$ should be
longer than the acceleration time $\tau_{ac} =
({3c^2/v_F^2})\tau_{sc}= 6\pi c/v_F^2 k_d = 3 c
v_A^4L/v_F^2u^3\min(v_F, u)\,,$ which implies $\max(u, v_F)
L/C_1u^2>cv_A^4L/v_F^2u^{3}\min(v_F, u)$, i.e., $ C_1<
u^2v_F^3/cv_A^4\,.$ There are at most two locations $x_1<x_2$ in the
down stream, where $\tau=\tau_{\rm ac}/\tau_{\rm d}=1$ and $C_1=
u^2v_F^3/cv_A^4$. In combination with equation (\ref{index}),
significant particle acceleration occurs for $p<\left[{9/4} +
{12v_F^2 \max(u^2, v_F^2)/u^4C_1^2}\right]^{1/2}-{1/ 2}\,.$ The
particle acceleration in the supersonic phase, i.e., $v_F<u$,
therefore produces very hard electron distributions with $p<1.39$
for $C_1=3$. Softer electron distributions have to be produced in
the subsonic phase. Figure \ref{paras} shows the evolution of
$\eta$, $\gamma_c$, $p$, and $\tau=\tau_{\rm ac}/\tau_{\rm d}$ in
the downstream for $U=0.01c$. The profiles of $v_F/U$ and $u/U$ only
depend on $v_A/U$. So is the profile of $\eta$. $\tau$ and $p$ also
depend on the absolute value of $U$. To obtain $\gamma_c$, one needs
to know $L$ and $B$ as well. In the extremely supersonic phase with
$v_F\ll u$, the SA is negligible. The SA is significant only after
the plasma is already heated up so that $v_F\sim u$. In the late
subsonic phase, $u\ll v_F$, the SA is also insignificant since most
of the free energy of the system has been converted into heat.

\begin{figure}[t]
\includegraphics[height=.27\textheight,width=0.48\textwidth]{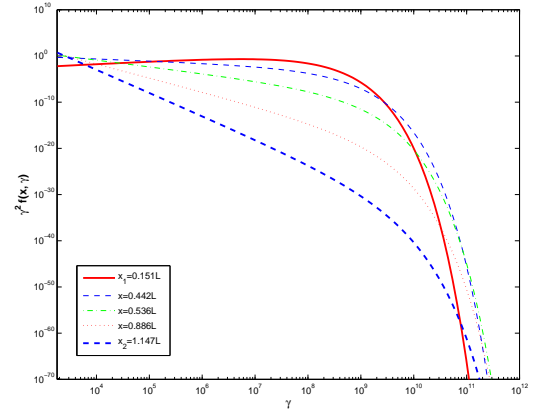}
\caption{Normalized nonthermal electron distribution $f$ produced at several
locations in the downstream. \label{f}}
\end{figure}

\begin{figure}[t]
\includegraphics[height=.27\textheight,width=0.48\textwidth]{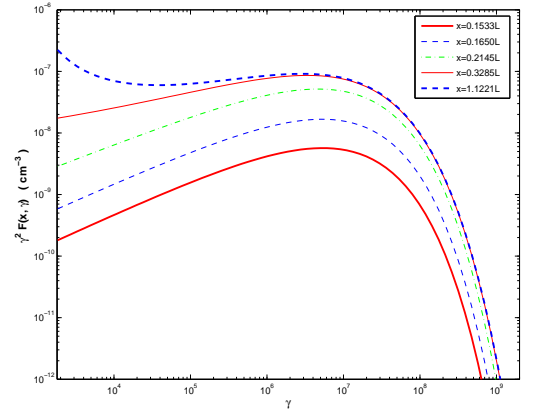}
\caption{The distributions of nonthermal electrons $F$ in the downstream. \label{Fy}}
\end{figure}

The particle distribution may be approximated reasonably well with
$f(x, \gamma)\propto \gamma^{-p(x)} \exp
-[\gamma/\gamma_c(x)]^{1/2}$\citep{l08}. Then the distribution of
non-thermal particle in the downstream
\begin{equation}\label{dis}
F(x, \gamma) = \int_{x_1}^x f(x^\prime, \gamma) \eta(x^\prime)
(4Q/m_ec^2U) {\rm d} x^\prime
\end{equation}
where $\int_{m_p/m_e}^\infty \gamma f(x^\prime, \gamma){\rm d}\gamma
= 1$, and $\int_{m_p/m_e}^\infty \gamma m_e c^2F(x, \gamma){\rm
d}\gamma$ gives the energy density of non-thermal particles at $x$.
If $u^5/cv_A^4<C_1$ at the sonic point $x_0$, then $x_0<x_1$ and
there will be no particle acceleration in the supersonic phase.
Figures \ref{f} and \ref{Fy} show the normalized electron
distribution $f$ and $F$ at several locations in the downstream,
respectively.

\section{Results}
\label{application}

\begin{figure}[t]
\includegraphics[height=.27\textheight,width=0.48\textwidth]{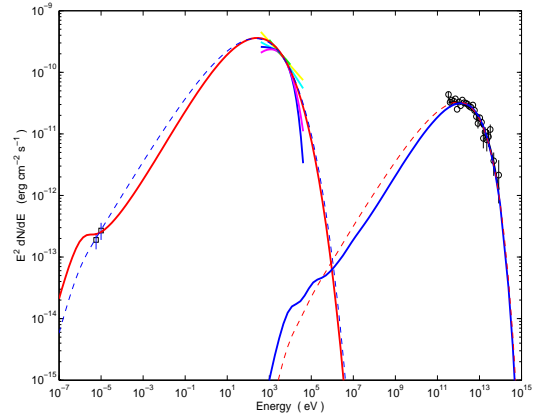}
\caption{Best fits to the observed spectrum. The dashed line is for
a simple power-law model with a gradual high energy cutoff. The
solid line is for the fiducial model. The low and high energy
spectral peaks are produced through the synchrotron and inverse
Compon scattering of the background photos, respectively.
\label{spec}}
\end{figure}

\begin{figure}[t]
\includegraphics[height=.27\textheight,width=0.48\textwidth]{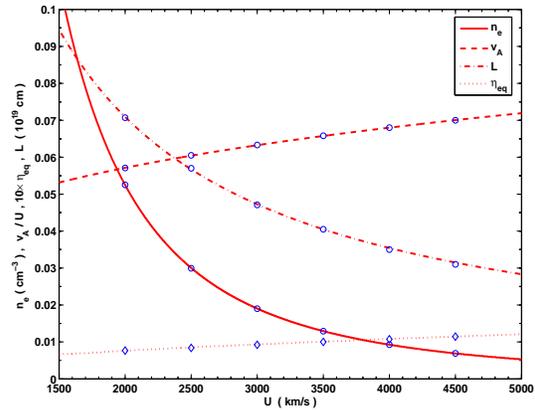}
\caption{Nearly identical fit to the observations are obtained for
parameters on these lines. \label{Udep}}
\end{figure}

Here, we use the SNR RX J1713.7-3946 as an example to demonstrate
how the SA by fast mode waves accounts for the observed broadband
spectrum. Figure \ref{spec} shows the best fit with $v_{\rm
A}/U=0.0633$, $L=4.71\times 10^{17}$ cm, $B=12$ $\mu$G and
$U=0.01c$. Comparing to the thin dashed line, which is derived by
assuming an electron distribution $\propto
\gamma^{-p}\exp-(\gamma/\gamma_c)^{1/2}$, there is a radio spectral
bump due to electron acceleration relatively far from the shock
front (see Figs. \ref{f} and \ref{Fy}). In our model, there are five
parameters: $B$, $U$, $v_{A}$, $L$, and the equivalent volume of a
uniform emission range. The last is 4 times bigger than the volume
of the SNR, suggesting higher nonthermal electron densities in the
interior of the remnant than near the shock front. The observed
radio to X-ray spectral index, X-ray to TeV flux ratio, location of
the X-ray cutoff, and bolometric luminosity of the source give four
constraints, which leads to one more degree of freedom. Our model
fit to the spectrum therefore is not unique. However, $B$ is
uniquely determined by the ratio of the X-ray to TeV flux. To
reproduce the observed spectral shape, the profiles of $p$,
$\gamma_c$, and $\eta$ should not change, which implies that
$v_A^8c^2\propto u^{10}$ and $L\propto u^4/v_A^4$ at the transonic
point $x_0$. For $v_A\ll U$, $u$ is proportional to $U$. We
therefore obtain $v_A$ and $L$ as functions of $U$ as indicated in
Figure \ref{Udep}. The density can be derived from $B$ and $v_A$,
and the overall acceleration efficiency is defined as $\eta_{eq} =
\int_{x_1}^{x_2}\eta Q {\rm d}x/\int_0^\infty Q{\rm d} x.$ Nearly
identical spectrum can be obtained for parameters on these lines. We
note $\eta\propto v_A^2/v_Sv_F\propto v_A^2/U^2\propto U^{1/2}$. The
acceleration is more efficient in the earlier phase of the remnant
evolution. The two thin dotted lines in Figure \ref{paras} show the
dependence of $\eta$ on $U$.

\end{document}